\documentclass[sn-nature]{sn-jnl}

\usepackage{graphicx}%
\usepackage{multirow}%
\usepackage{amsmath,amssymb,amsfonts}%
\usepackage{amsthm}%
\usepackage{mathrsfs}%
\usepackage[title]{appendix}%
\usepackage{xcolor}%
\usepackage{textcomp}%
\usepackage{manyfoot}%
\usepackage{booktabs}%
\usepackage{algorithm}%
\usepackage{algorithmicx}%
\usepackage{algpseudocode}%
\usepackage{listings}%

\theoremstyle{thmstyleone}%
\theoremstyle{thmstyletwo}%

\theoremstyle{thmstylethree}%

\begin{document}

\title{Comparison between neural network clustering, hierarchical clustering and k-means clustering: Applications using fluidic lenses}


\author[1,2]{ Graciana Puentes}


\affil[1]{Departamento de Fisica, Facultad de Ciencias Exactas y Naturales, Universidad de Buenos Aires, Ciudad Universitaria, 1428 Buenos Aires, Argentina}

\affil[2]{CONICET-Universidad de Buenos Aires, Instituto de Fisica de Buenos Aires (IFIBA), Ciudad Universitaria, 1428
Buenos Aires, Argentina}


\abstract{A comparison between neural network clustering (NNC), hierarchical clustering (HC)  and K-means clustering (KMC) is performed to evaluate the computational superiority of these three machine learning (ML) techniques for organizing large datasets into clusters. For NNC, a self-organizing map (SOM) training was applied to a collection of wavefront sensor reconstructions, decomposed in terms of 15 Zernike coefficients, characterizing the optical aberrations of the phase front transmitted by fluidic lenses. In order to understand the distribution and structure of the 15 Zernike variables within an input space, SOM-neighboring weight distances, SOM-sample hits, SOM-weight positions and SOM-weight planes were analyzed to form a visual interpretation of the system's structural properties. In the case of HC, the data was partitioned using a combined dissimilarity-linkage matrix computation. The effectiveness of this method was confirmed by a high cophenetic correlation coefficient value (c=0.9651). Additionally, a maximum number of clusters was established by setting an inconsistency cutoff of 0.8, yielding a total of 7 clusters for system segmentation. Nevetheless, we found this result is dramatically modified by setting a different inconsistency cutoff, resulting in the formation of 13  different clusters, when reducing the cutoff from 0.8 to 0.7. In addition, a KMC approach was employed to establish a quantitative measure of clustering segmentation efficiency, obtaining a sillhoute average value of 0.905 for data segmentation into K=5 non-overlapping clusters. On the other hand,  the NNC analysis revealed that the 15 variables could be characterized through the collective influence of 8 clusters.  It was established that the formation of clusters through the combined linkage and dissimilarity algorithms of HC alongside KMC is a more dependable clustering solution than separate assessment via NNC or HC, where altering the SOM size or inconsistency cutoff can lead to completely new clustering configurations. 
}

\maketitle

\section{Introduction}

Clustering refers to the method of grouping variables with similar characteristics into sets known as clusters. Given the importance of clustering in areas such as data analysis, pattern recognition, image processing, information retrieval, and medical imaging, it has been extensively researched using a variety of computational techniques. An effective clustering algorithm proficiently manages tasks such as scaling non-uniform data, analysing categorical, numerical, and binary data, interpreting results thoroughly, uncovering key features of a system, and dealing with high-dimensional data. The most frequently utilized clustering techniques include Hierarchical Clustering (HC) and K-Means Clustering (KMC). In HC, a hierarchy of clusters is formed, with each cluster comprising data points that share similar characteristics. KMC, on the other hand, creates clusters by averaging the data around K centroids. Zhao et al. \cite{Zhao} noted that the consistency of HC solutions across various levels of detail enabled flat partitions of differing granularity, making them well-suited for interactive exploration and visualization. In scenarios where clusters contain numerous sub-clusters, a hierarchical structure was considered a natural constraint within the relevant application domain, such as biological taxonomy and phylogenetic trees. Recently, in the pursuit of effective clustering algorithms to articulate the learned representation necessary for feature detection, Neural Network Clustering (NNC) has gained preference.\\

Cirrincione et al. \cite{Cirrincione} suggested that the limitation of threshold splitting in HC could be resolved through NNC by constructing a hierarchical tree in a progressive and self-organizing manner. This approach, based on the innovative concept of neighborhood convex hulls, characterizes horizontal expansion through an anisotropic influence region. Additionally, the combination of hierarchical segmentation and the GH-EXIN neural network was seen as a means to enhance clustering accuracy. Zhang et al. \cite{Zhang} demonstrated that combining hierarchical dendrograms with heat maps offered a clear visualization of clinical studies involving diverse study populations. A system known as Cognitive Comparison-Enhanced Hierarchical Clustering (CCEHC) was introduced by Guan, C. and Yuen \cite{Guan} to deliver tailored product suggestions based on individual user preferences. Du \cite{Du} emphasized the significance of self-organizing maps (SOM) as a competitive learning-based clustering neural network, capable of extracting valuable information from vast databases or the World Wide Web (WWW). It was highlighted that structural features must first be identified in a database for data mining, and the self-organization exploratory technique appeared particularly promising. Jain et al. \cite{Jain} identified database segmentation, predictive modelling, and visualization of large databases as some clustering strategies for data mining. It was argued that web mining presents challenges due to the ambiguous characteristics of the less structured WWW database. In this context, the topology-preserving attribute of NNC rendered it especially appropriate for processing visual web information.\\

Mangiameli et al. \cite{Mangiameli} performed a comparison between NNC and HC, demonstrating that NNC outperformed HC. It was indicated that HC methods were prone to making classification mistakes when real-world data deviated from the ideal conditions of tightly grouped isolated clusters. However, the advantage of NNC over HC was firmly reliant on achieving precise decision-making through NNC on partially organized data. It is worth mentioning that Refs. \cite{Cirrincione, Mangiameli, Herrero, Okamoto} underscored the enhancement of clustering accuracy via an NNC-HC combination. Recently, Shahid \cite{Shahid} reported a comparison between NNC and HC precision dominance, applied to  a local aqua system to learn the distribution and topology of variables in an input space using a total of 8 neurons, resulting in the formation of 6 clusters for NNC. They concluded that HC provides for a more dominantly precise clustering characterisation than NNC, resulting in 7 clusters for data segmentation considering an inconsistency cutoff of 0.7842. \\

In this paper, we present a comparison between NNC, HC and KMC applied to a completely different scenario. Namely, a set of 15 Zernike coefficients characterising optical aberrations obtained by wavefront reconstruction of the phase front transmitted by fluidic lenses of varying fluidic volumes using a Shack-Hartmann wavefront sensor \cite{1,2,3,4,5,6,7,8,9,10,10b,11,12,13,14,15,16,17,18,19,20,21,22,23}. Moreover, as opposed to  Ref. \cite{Shahid}, we construct a NN using 10 neurons, resulting in the formation of 8 clusters for system segmentation. We found that HC can provide for a more precise clustering configuration consisting of 7 clusters, as long as a sufficiently high inconsistency cutoff is chosen ($ \leq 0.8$). In contrast to Ref. \cite{Shahid}, we encounter that when the inconsistency cutoff is reduced from 0.8 to 0.7, the clustering configuration changes dramatically, resulting in the formation of 13 new clusters, clearly reducing the precision of HC in comparison with NNC. Finally, a KMC approach was employed to establish a quantitative measure of clustering segmentation efficiency, obtaining a silhouette mean value of 0.905 for data segmentation into $K=5$ clusters. We conclude that while NNC characterization is mostly visual, and HC precision dominance is strongly dependent on the elected inconsistency cutoff, KMC can provide an efficient clustering characterization, as quantified by the mean silhouette value. Thus NNC, HC and KMC,  can be considered as complementary techniques, which should be employed jointly for a more dependable clustering characterization. \\

\subsection{Data Components}

The data used for the current clustering analysis is comprised of 15 Zernike coefficients for 8 different weights, obtained by using a Shack-Hartmann Wavefront Sensor to reconstruct the wavefront transmitted by fluidic lenses of varying fluidic volumes \cite{1,2,3,4,5,6,7,8,9,10,11,12,13,14,15,16,17,18,19,20,21,22,23}, as depicted in Figure 1 (a). The 15 Zernike coeffecients ($Z_1-Z_{15}$) and their associated Zernike polynomial for characterisation of optical aberrations are depicted in Table 1 left column and right column, respectively. The number assigned to the Zernike variables depicted in Table 1 (middle column) correspond to the numbering system used for HC and KMC analysis. 


\begin{table}[h]
\begin{tabular}{@{}llll@{}}
\toprule
\textbf{Zernike coefficient}  & \textbf{Nr Assigned to Variable} & \textbf{Zernike polynomial}  \\
\midrule
$Z_{0}$ & 1 &1 \\
$Z_{1}$ & 2 & $2\xi \sin \phi$  \\
$Z_{2}$ & 3 & $2\xi \cos \phi$ \\
$Z_{3}$ & 4 & $\sqrt{6}\xi ^{2}\sin 2\phi $ \\
$Z_{4}$ & 5 & $\sqrt{3}\left( 2\xi ^{2}-1\right) $ \\
$Z_{5}$ & 6 &$ \sqrt{6}\xi ^{2}\cos 2\phi $\\
$Z_{6}$ & 7 & $\sqrt{8}\xi ^{3}\sin 3\phi$\\
$Z_{7}$ & 8 &$ \sqrt{8}\left( 3\xi ^{3}-2\xi \right) \sin \phi $\\
$Z_{8}$ & 9 & $\sqrt{8}\left( 3\xi ^{3}-2\xi \right) \cos \phi $\\
$Z_{9}$ & 10 &$ \sqrt{8}\xi ^{3}\cos 3\phi$ \\
$Z_{10}$ & 11 & $\sqrt{10}\xi ^{4}\sin 4\phi $\\
$Z_{11}$ & 12 &$ \sqrt{10}\left( 4\xi ^{4}-3\xi ^{2}\right) \sin 2\phi $ \\
$Z_{12}$ & 13 &$ \sqrt{5}\left( 6\xi ^{4}-6\xi ^{2}+1\right) $\\
$Z_{13}$ & 14 & $\sqrt{10}\left( 4\xi ^{4}-3\xi ^{2}\right) \cos 2\phi $ \\
$Z_{14}$ & 15 & $\sqrt{10}\xi ^{4}\cos 4\phi $\\
\botrule
\end{tabular}
\caption{\label{tab:widgets}Left Column: Zernike coefficients obtained by wavefront sensing ($Z_1-Z_{15}$). Right column: Associated Zernike polynomial. Middle column: Number Assigned to Variables $Z_1-Z_{15}$ for HC and KMC analysis. }
\end{table}

\begin{figure}
\includegraphics[width=1.0\linewidth]{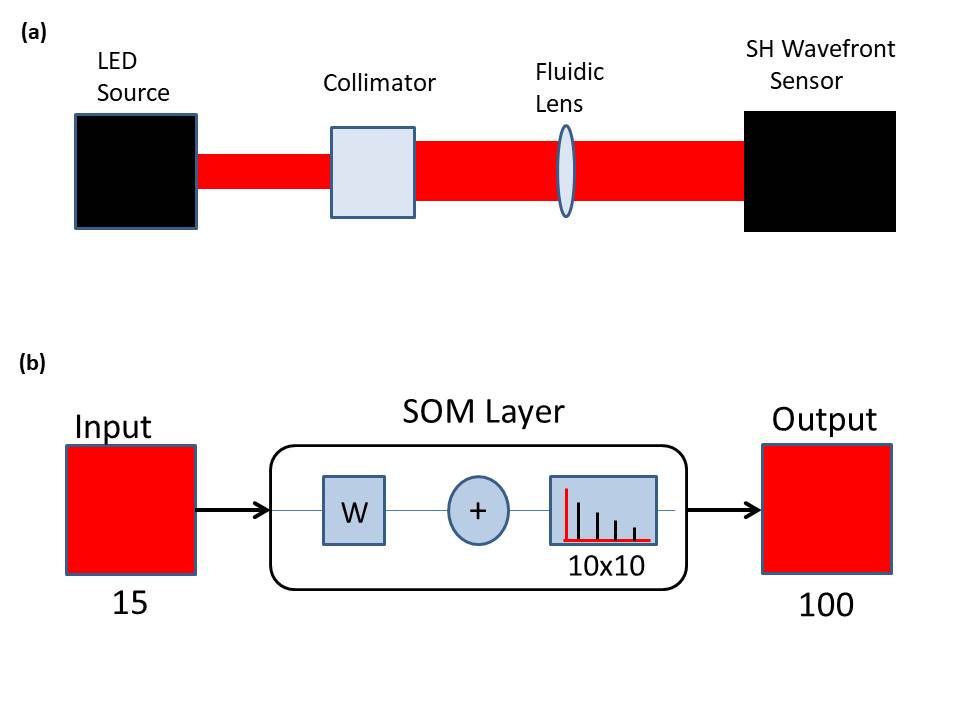}
\caption{\label{fig:frog} (a) Experimental setup for reconstruction of Zernike coefficients characterizing the phase front transmitted by fluidic lenses of different fluidic volumes. (b) Training network for 15 Input variables ($Z_1-Z_{15}$), considering 8 weights and a neural network with 10 neurons. }
\end{figure}

\section{Statistical Methods}

In order to identify a more effective machine learning (ML) technique that offers an optimized analysis of pattern-based outcomes from extensive data, three approaches are utilised: Neural Network Clustering (NNC),  Hierarchical Clustering (HC), and K-Means Clustering (KMC). The characteristics emphasized by each method are evaluated to illustrate which of the strategies excels in delivering precision and efficiency.

\subsection{Neural Network Training}

For an input of 15 Zernike variables, a neural network (NN) consisting of 10 neurons, as shown in Fig. 1(b), was trained until it reached the stopping criteria of 200 epochs of the ML Batch algrorithm. The dimensions of the Self-Organizing-Map (SOM)  corresponds to the number of neurons illustrated on the horizontal axis in Fig. 2. In Fig. 2, data comprising 8 predictor variables is input into an SOM network, allowing neurons to layer and group based on shared characteristics. Ultimately, the SOM organizes the related neurons into a 100-grid map.

\subsubsection{SOM Neighbor Weight Distances}

To represent all weights within an 8-dimensional input space—consisting of 8 different fluidic volumen samples for each of the 15 input Zernike variables—we create a plot of SOM neighbor weight distances, illustrated in Fig. 2. This plot is characterized by a hexagonal grid known as SOM topology, where each blue hexagon corresponds to a neuron. Throughout the training phase, the weight vector linked to each neuron adjusts to become the center of a cluster of input vectors. The attributes of the plot are clarified using color-coding: blue hexagons signify the neurons (cluster centers/weight vectors), while red lines connect neighboring neurons. The distances among neurons are depicted through colors in the areas containing the red lines; darker colors indicate larger distances, whereas lighter colors signify smaller distances. A total of 8 clusters can be identified based on the yellow coloured patches limited by darker (orange, red, black) borders. 

\begin{figure}
\includegraphics[width=1\linewidth]{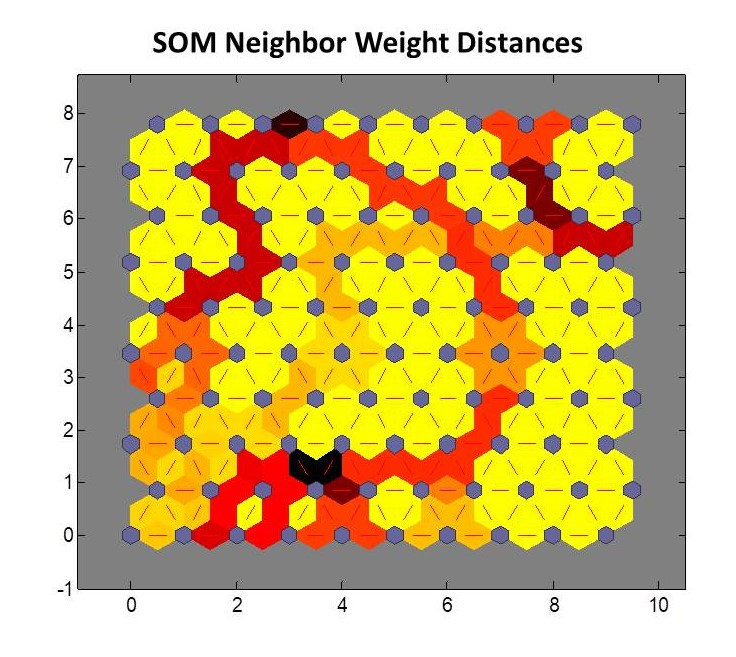}
\caption{\label{fig:frog} Self-Organizing Map (SOM) Neighbor Weight Distances depicting SOM Topology. Horizontal axis indicates the number of neurons.  }
\end{figure}

\subsubsection{SOM Sample Hits}

To evaluate the number of input vectors linked to each neuron, we generate an SOM sample hits plot (see Fig. 3). The number displayed on a shaded neuron indicates its association with an input vector. This visualization reveals that the highest number of input vectors associated with a single neuron is 1.  It is best if the data are fairly evenly distributed across the neurons. In this example, the data are concentrated a little more in the lower left corner neurons.

\begin{figure}
\includegraphics[width=1\linewidth]{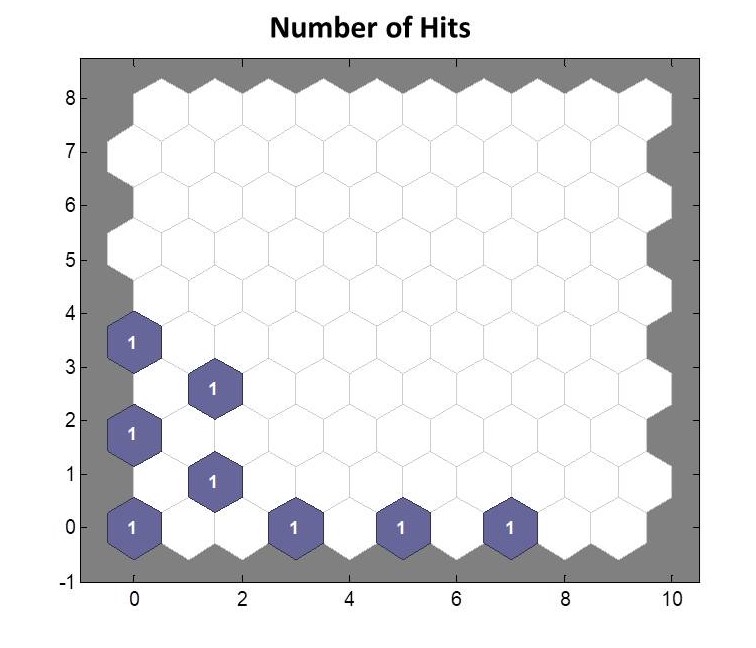}
\caption{\label{fig:frog} SOM Sample Hits depicting the number of input vectors linked to each neuron.}
\end{figure}

\subsubsection{SOM Input Planes}

To visualise the weights associated with the elements of input vectors, a weight plane figure is generated. Each element of an input vector is represented by its corresponding weight plane. Since there are 15 input vectors ($Z_1-Z_{15}$), each containing 8 elements, a weight plane is produced for each element. These weight planes illustrate the connections between individual input elements and the neurons in the network. In the plot, lighter colors indicate larger weights, while darker colors signify smaller weights. When input elements exhibit similar connection patterns, it reflects a strong correlation among those elements.\\

\noindent Figure 4 illustrates subplots for various input Zernike coefficients ranging from $Z_1$ to $Z_{15}$. In each plot, the connections between weights of specific inputs and the layer’s neurons are depicted using three distinct colors. Yellow represents the strongest positive connections, red indicates no connection, and black signifies the strongest negative connections. Key observations from Figure 4 are as  follows:\\

\begin{itemize}

\item{Based on the position of the yellow neuron cluster on top right corner  of the planes a high correlation between $Z_1$, $Z_7$, and $Z_{13}$ is indicated.}

\item{Visual examination of the position of the black neuron cluster on top right plane corner indicates high correlation between $Z_2$, $Z_3$, and $Z_9$.}

\item{Viewing the positions of yellow neuron cluster on top left corner of the planes indicates high correlation between $Z_2, Z_5, Z_6, Z_{14}, Z_{15}$.}

\item{Based on the position of black neuron cluster on top left corner of the planes a high correlation between $Z_1, Z_8, Z_{11}, Z_{12}$ is indicated.}

\item{Visual examination of black neuron cluster in the middle of the plane indicates high correlation between $Z_{10}, Z_{13}, Z_{14}$.}

\item{Viewing the positions of the orange neuron cluster in the middle of the planse indicates high correlations between $Z_3, Z_7, Z_{11}$.}

\item{Visual inspection of red neuron clusters in the middle of the planes indicates high correlation between $Z_2, Z_5, Z_{15}$.}

\end{itemize}

\begin{figure}
\hspace{-7cm}
\includegraphics[width=2\textwidth]{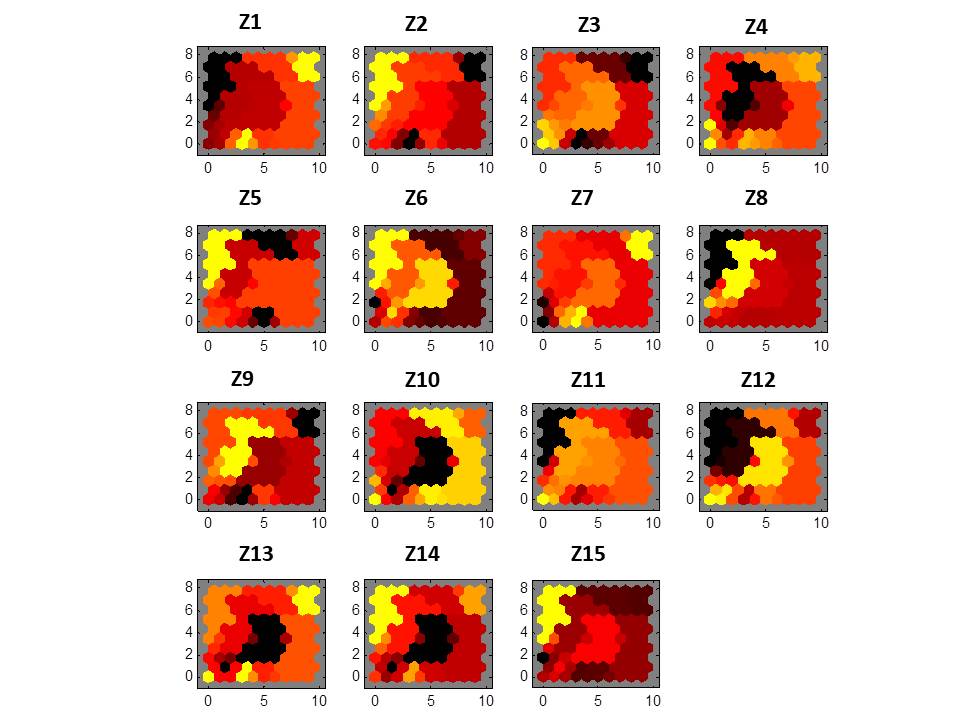}
\caption{\label{fig:frog} SOM Input Planes depicting strong correlations between different input vectors ($Z_1-Z_{15}$).}
\end{figure}

\subsubsection{SOM Weight Planes}

The SOM Weight planes  shows the locations of the data points and the weight vectors. As Figure 5 indicates, after 200 iterations of the ML Batch algorithm, the map is distributed through the input space. SOMs differ from conventional competitive learning in terms of which neurons get their weights updated. Instead of updating only the winner, feature maps update the weights of the winner and its neighbors. The result is that neighboring neurons tend to have similar weight vectors, and to be responsive to similar input vectors. The figure indicates the following color coding: \\

\begin{itemize}
\item{Blue hexagons indicate neurons.}

\item{The red lines connect neighboring neurons.}

\item{The colors in the regions containing the red lines indicate the distances between neurons.}

\item{The darker colors represent larger distances.}

\item{The lighter colors represent smaller distances.}

\end{itemize}

\begin{figure}
\includegraphics[width=1\linewidth]{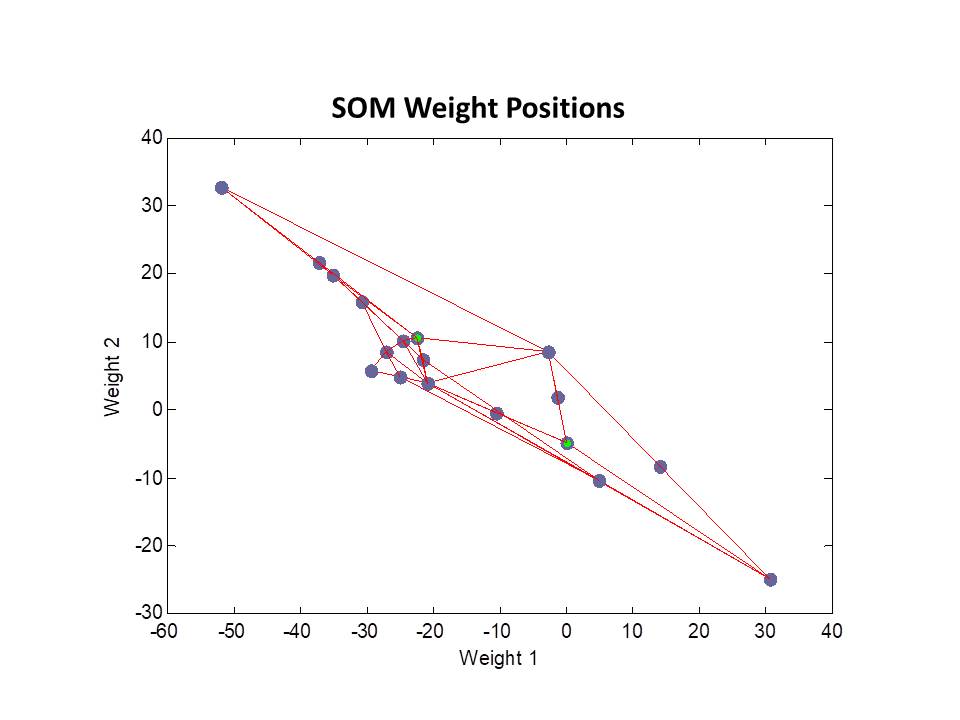}
\caption{\label{fig:frog} SOM Weight Planes displaying the locations of data points and the weight vectors.}
\end{figure}

\subsection{Hierachical Clustering}

\begin{figure}
\includegraphics[width=1\linewidth]{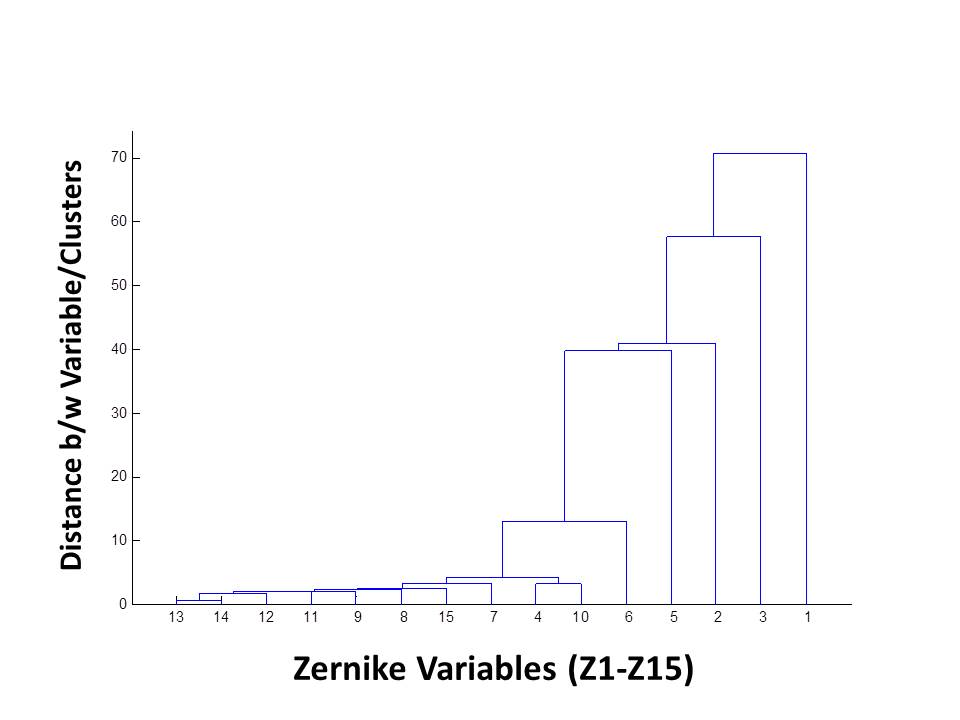}
\caption{\label{fig:frog} Dendrogam for clusters. }
\end{figure}

To identify variables with similar attributes in Zernike coefficient data, a Hierarchical Clustering \cite{37,38} method was employed. This approach  groups data over a variety of scales by creating a cluster tree or dendrogram. The tree is not a single set of clusters, but rather a multilevel hierarchy, where clusters at one level are joined as clusters at the next level. This allows to decide the level or scale of clustering that is most appropriate, enabling the analysis of integrated movement or the influence of specific clusters within a system. The HC process was applied to the 15 Zernike input variables. The numbers assigned to the variables for the HC analisis is depicted in Table 1 (middle column).\\


\begin{table}[h]
\begin{tabular}{@{}llllllllllllllll@{}}
\toprule
\hspace{-2.4cm} \textbf{Sr} &1 & 2 & 3 & 4 & 5 & 6 & 7 & 8 & 9 &10 & 11 & 12 & 13 & 14 &15\\
\midrule
\hspace{-2.4cm}\textbf{Nr. of links} & 1 & 2 & 2 & 2 & 2 &2 & 1 & 2 & 3 & 2 & 2 & 2 & 2 & 2 &2 \\
\hspace{-2.4cm}\textbf{Inconsistency } & 0 & $0.707 $ & 0.707 & 0.707 & 0.707 &0.707 & 0 & 0.707 & 1.153 & 0.707 & 0.707  & 0.707 & 0.707  & 0.707& 0.707 \\
\botrule
\end{tabular}
\caption{\label{tab:widgets}Inconsistency measure for links in Dendrogram.}
\end{table}

\subsubsection{Dendrogram (cluster tree)}

\begin{figure}
\includegraphics[width=1\linewidth]{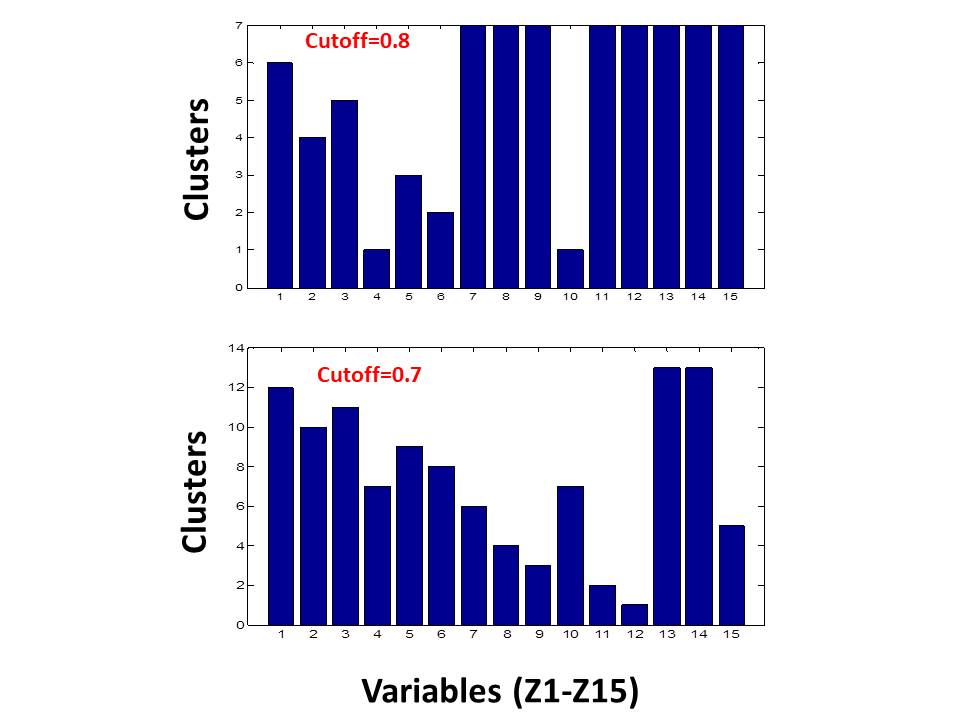}

\caption{\label{fig:frog} (a) Bar plot for inconsistency cutoff=0.8, revealing the creation of 7 clusters. (b) Bar plot for inconsistency cutoff=0.7, revealing the creation of 13 new clusters.}
\end{figure}

To organize 15 Zernike variables into groups, a cluster tree was generated using the Machine Learning (ML) program "clusterdata." This tree represents hierarchical stacks of clusters across different levels, each level grouping variables that demonstrate a similar influence on the overall system. Applying the clusterdata function to the normalized values of a 15-variable dataset helped construct a dendrogram (Fig. 6) using a dissimilarity matrix and a linkage matrix. The dissimilarity matrix quantified the distance between each pair of variables, while the linkage matrix identified connections between variable pairs or clusters. Moreover, the linkage function calculated distances not only between individual variables but also between clusters or between a variable and a cluster, enhancing the clustering analysis. 

\subsubsection{Cophenetic Coefficient }

To verify the creation and connection of clusters displayed in the cluster tree on Fig. 6, the cophenetic coefficient is computed, obtaining a numerical value $c=0.9651$. This metric measures the correlation between the distance matrix and the linkage matrix. A higher coefficient value, closer to 1, indicates that the distances between clusters or variables joined by the linkage function align well with the distances between variables determined by the dissimilarity function. The cophenetic correlation coefficient for the dendrogram confirms the proximity of linked variables or clusters in relation to their true distances within the input space.

\subsubsection{Inconsistency Measure for links}

To partition a set of variables into natural clusters, an inconsistency measure is calculated. The smallest value of this measure between any two variables or clusters indicates close resemblance or high similarity between them. The inconsistency measure serves to distinguish clusters based on the heights of the links connecting them. Table 2 provides the inconsistency measures for the links in the Dendrogram shown in Fig. 5. The second row lists the number of links across different levels, while the third row displays their respective inconsistency coefficients. An inconsistency coefficient of '0' corresponds to the links connecting variable pairs 4, 10, 13 and 14 as depicted in Fig. 6, since no links exist below these variables, in good agreement with the correlations observed in the SOM-Input planes figure.  

\begin{figure}
\includegraphics[width=1\linewidth]{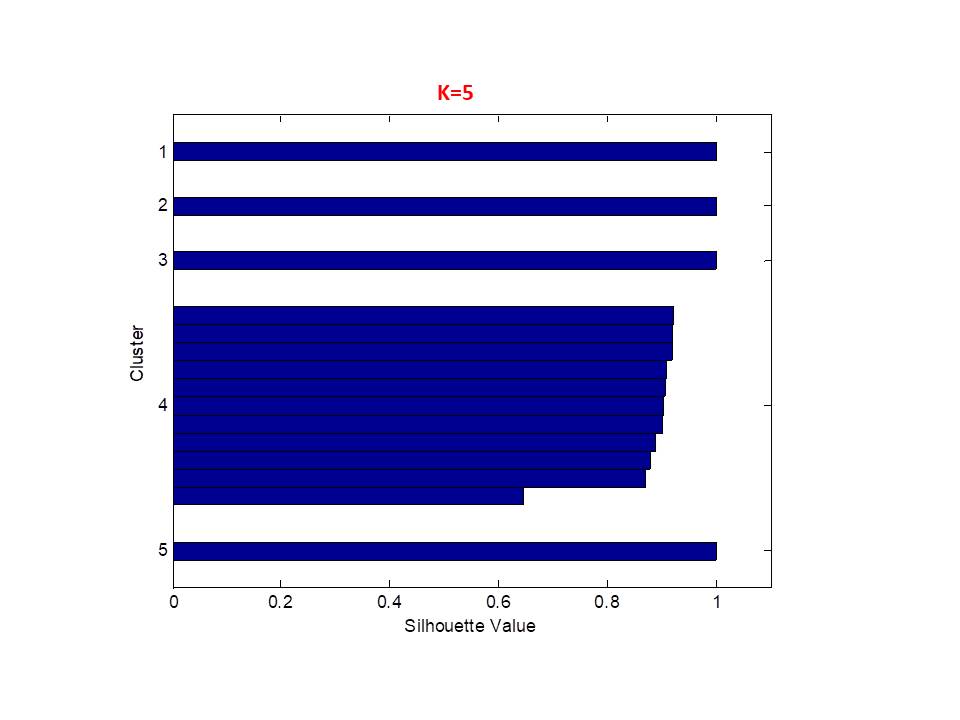}

\caption{\label{fig:frog} Silhouet plot for $K=5$.}
\end{figure}

\begin{figure}
\includegraphics[width=1\linewidth]{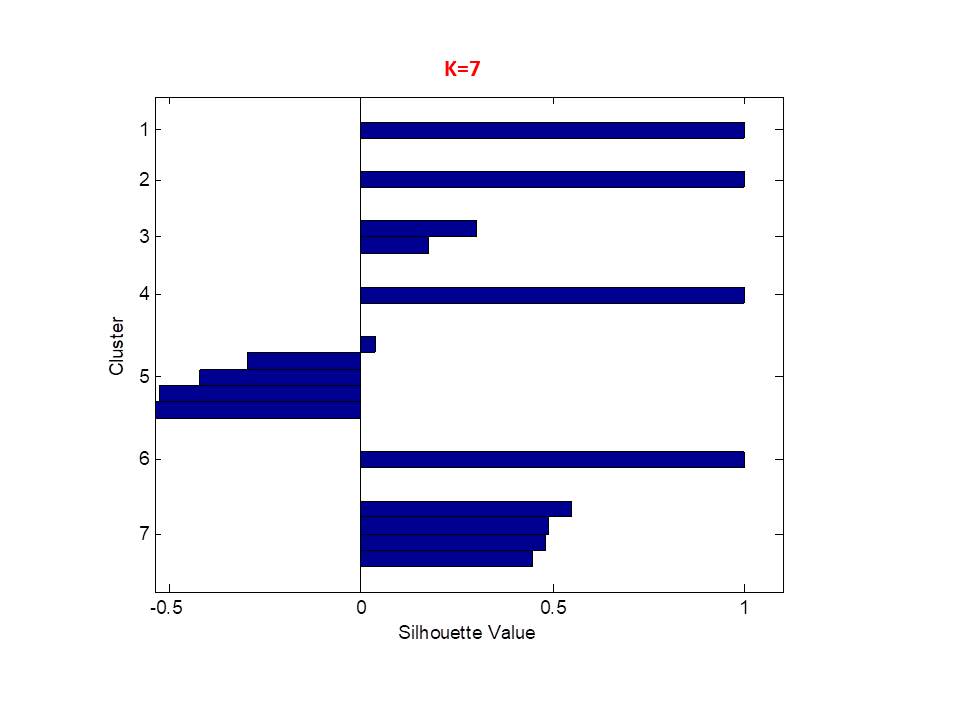}

\caption{\label{fig:frog} Silhoute plot for $K=7$}
\end{figure}

\subsubsection{Bar plots}

The hierarchical cluster tree can naturally divide the data into distinct, well-separated clusters. This can be particularly evident in a dendrogram diagram created from data where groups of objects are densely packed in certain areas and not in others. The inconsistency coefficient of the links in the cluster tree can identify these divisions where the similarities between objects change abruptly. This inconsistency value can be used to determine where the cluster function creates cluster boundaries. To quantify this concept we use the *cluster* function to group the sample data set into clusters, specifying an inconsistency coefficient threshold of 0.8 as the value of the cutoff argument, the cluster function groups all the objects in the sample data set into 7 cluster distinct clusters, as depicted in the bar plot Figure 7 (a). Note, that this clustering configuration can be changed dramatically by setting a different inconsistency cutoff. For example, in Figure 7 (b) we display the bar plot reducing the inconsistency cutoff from 0.8 to 0.7. A completely different clustering configuration is found, consisting of 13 new clusters. This feature evidences that the cluster configuration obtained by HC is strongly dependent on the inconsistency cutoff elected, displaying entirely new clusters when varying the cutoff from 0.8 to 0.7.  

\subsection{K-Means Clustering}

A third Machine Learning (ML) clustering method, namely K-Means Clustering (KMC), was employed to compare its efficiency with respect to NNC and HC approaches. K-Means clustering is a partitioning method determined by the ML function *kmeans* which partitions the data into K mutually exclusive clusters and returns the index of the cluster to which it assigns each observation. K-Means treats each data point as an object that has a location in space. The function finds a partition in which objects within each cluster are as close to each other as possible, and as far from objects in other clusters as possible. In our application we selected the distance metric "cityblock" (sum of absolute differences, i.e., the L1 distance) although other distance metrics are also possible. By default, kmeans begins the clustering process using a randomly selected set of initial centroid locations. The kmeans algorithm can converge to a solution that is a local (nonglobal) minimum; that is, kmeans can partition the data such that moving any single point to a different cluster increases the total sum of distances. However, as with many other types of numerical minimizations, the solution that kmeans reaches sometimes depends on the starting points. Therefore, other solutions (local minima) that have lower total sums of distances can exist for the data. Unlike HC, K-Means clustering operates on actual observations rather than on the dissimilarity between pairs of observations in the data. Also, K-Means clustering creates a single level of clusters, rather than a multilevel hierarchy of clusters, such as HC. Therefore, K-Means clustering is often more suitable than HC for large amounts of data. Each cluster in a K-Means partition consists of member objects and a centroid. In each cluster, K-Means minimizes the sum of the distances between the centroid and all member objects of the cluster. The numbers assigned to the Zernike variables used in the KMC analysis is depicted in Table 1 (middle column). 

\subsubsection{Silhouette Plots}

A silhouette plot displays a measure of how close each point in one cluster is to points in the neighboring clusters. This measure ranges from 1 (indicating points that are very distant from neighboring clusters) through 0 (points that are not distinctly in one cluster or another) to –1 (points that are probably assigned to the wrong cluster). Figure 8 shows the sillhouette plot setting to total number of mutually exclusive clusters to $K=5$. The silhouette plot indicates that the 5 clusters are very well separated, since most sillhoutte values assigned to the 15  Zernike variables ($Z_1-Z_{15}$) are higher than 0.8, with no low or negative values assigned to any of the variables. On the other hand, figure 9 shows the sillhouette plot setting the total number of mutually exclusive clusters to $K=7$. Figure 9 indicates that  7 clusters is a less efficient clustering segmentation, since there are many low and negative values indicating points not distinctly classified, or merely assigned to the wrong cluster. A more quantitative approach to comparing clustering solutions can be obtained by computing the mean silhouette values for the two cases. For $K=5$ we obtain an average sillhoutte  value of 0.9105 while for $K=7$ we obtain an avarage sillhoutte value of 0.3138.  An average value closer to 1 indicates a more efficient clustering classification. Thus we can safely conclude that 5 mutually exclusive clusters ($K=5$) is the most efficient clustering segmentation for the data points analyzed. \\


\begin{table}[h]
\begin{tabular}{@{}llll@{}}
\toprule
\textbf{K}  & \textbf{Mean} & \textbf{Sum Distance} \\
\midrule
5 & 0.9105 & 82.402 (4 it)\\
7 & 0.3138 & 62.872 (2 it) \\
\botrule
\end{tabular}
\caption{\label{tab:widgets}Left Column: Numer of clusters ($K$). Middle column: Mean sillhoute value. Right column: Sum of total distances. }

\end{table}

On Table 3 we present the quantitative results obtained by K-Means clustering (KMC) for different number of mutually exclusive clusters ($K$). Namely, for $K=5$ we obtain a mean sillhouette value of 0.9105, and a total sum of distances of 82.402, obtained after 4 iterations of the ML algorithm. On the other hand, for $K=7$ we obtain a mean sillhoutte values of 0.3138, and a total sum of distances of 62.872 obtained after 2 iterations of the ML algorithm.  As expected the sum of total distances decreases as $K$ increases. We can safely conclude that $K=5$ is a more efficient clustering segmentation since a mean sillhoute value closer to one indicates a more efficient partitioning. 

\section{Discussion}

A study was carried out to compare K-means clustering (KMC), hierarchical clustering (HC) and neural network clustering (NNC) in dividing data into clusters. HC offers a systematic computational approach to separate system components based on specific characteristic counts of the predictor variables. The resulting partitions were further optimized using metrics such as the cophenetic coefficient and inconsistency coefficient. Similarly, KMC was utilized to establish an optimal clustering configuration as quantified by the mean sillhoutte value. In addition, NNC was employed to highlight visual distinctions and similarities among predictor variables, leveraging training plots for enhanced clarity in the analysis.\\

A Self-Organizing Map (SOM)-Neural Network Clustering was utilized to organize Zernike variables with similar characteristics into distinct clusters. A Machine Learning (ML) Algorithm facilitated the classification of variables by mapping them into an input space encompassing 15 locations and 8 parametric values. Three key aspects of the SOM—Sample Hits, Neighboring Weight Distances, Weight Planes and Weight Positions—were analyzed to assess the distribution and consolidation of $Z_1-Z_{15}$  Zernike components. The visualization of sample hits suggests that the 16 sampled locations can be categorized into eight distinct clusters based on data conformity (Fig. 3). In addition to this, the dissimilarity among the 16 vectors was assessed through Neighbour Weight Distances as shown in Fig. 2. Approximately 8 clusters can be identified according to high correlation, indicated by yellow areas surrounded by dark red or black patches in the maps. To identify highly correlated Zernike variables, a weight plane visualization (Fig. 4) was generated to display the weights associated with each parameter for individual neurons. Based on similar patterns across the weight planes and clustering of larger weights near lighter-colored neurons, 8 distinct clusters were identified. We note the number of clusters identified strongly depends on the number of neurons and input weigths elected. Varying SOM size can lead to a completely different clustering configuration. \\

A Machine Learning (ML) algorithm, *clusterdata*, was utilized to generate a Dendrogram cluster tree, as depicted in Fig. 6. This tree illustrates different hierarchical levels, each containing clusters of Zernike variables. The integration of a dissimilarity function and a linkage function enabled the pairing of clusters or the association of a variable with a cluster based on their closest proximity distances. To determine whether the grouping of variables into clusters within the dendrogram accurately represents their similarity or divergence in a real system, a cophenetic coefficient was calculated, resulting in $c=0.9651$. A higher cophenetic value (closer to 1) confirms the dendrogram's effectiveness in clustering data based on dissimilarity features. Additionally, a cutoff inconsistency coefficient, which compares the link heights within the dendrogram, was employed to identify the exact number of clusters encompassing 15 variables. The classification process, illustrated in Fig. 7, organized the variables into 7 clusters for an inconsistency cutoff of 0.8, and 13 clusters for an inconsistency cutoff of 0.7,  based on the similarity of their link heights, ensuring that variables with comparable link lengths were grouped together. This revealed that the clustering configuration is extremely sensitive to the elected cutoff. \\

Finally the ML algorithm *kmeans* was utilized to compare the efficiency of K-means clustering (KMC) with respect to NNC and HC approaches. By constructing sillhoutte plots for different mutually exclusive clustering numbers $K$ (Figures 8 and 9), and computing the average sillhoute value, it was quantitatively established that the mosts efficient clustering segmentation results for $K=5$ clusters, characterized by an average sillhoutte value of 0.9105, while clustering partitioning for $K=7$ results in a significantly lower average sillhoute value of 0.3138, with several negative sillhouette values, indicating wrong clustering assignments. The current analysis reveals that while NNC provides for a clear visual inspection of the resulting clusters, KMC can provide for an efficient clustering segmentation as quantified by a high average sillhoutte value. On the other hand, HC can provide for a more accurate hierarchical clutering characterization, as quantified by a high correlation cophenetic coefficient. Nevertheless, while NNC partitioning results vary significantly for different SOM sizes, HC depends strongly on the elected inconsistency threshold cutoff.  This implies that the three different ML clustering approaches can be considered complementary, and should be employed jointly for a more dependable clustering classification. \\

\section{Acknowledgements} 
The author acknowledges financial support from PICT Startup 2015 0710 and Raices Programme.


\begin{thebibliography}{}

\bibitem{Zhao} Zhao, Y., Karypis, G.  Fayyad, and U. Hierarchical clustering algorithms for document datasets. Data Min. Knowl. Disc. \textbf{10}, 141–168 (2005).

\bibitem{Cirrincione} Cirrincione, G., Ciravegna, G., Barbiero, P., Randazzo, and V.  Pasero, E. The GH-EXIN neural network for hierarchical clustering. Neural Netw. \textbf{121}, 57–73.

\bibitem{Zhang} Zhang, Z., Murtagh, F., Van-Poucke, S., Lin, S. and  Lan, P. Hierarchical cluster analysis in clinical research with heterogeneous study population: Highlighting its visualization with R. Ann. Transl. Med. \textbf{5}, 75 (2017).

\bibitem{Guan} Guan, C.  Yuen, K. K. F. The cognitive comparison enhanced hierarchical clustering. Granul. Comput. \textbf{7}, 637–655 (2022).

\bibitem{Du} Du, K.-L. Clustering: A neural network approach. Neural Netw. \textbf{23}, 89–107 (2010).

\bibitem{Jain} Jain, A., Murty, M. and Flynn, P. Data clustering: A review. ACM Comput. Surv. \textbf{31}, 264–323 (1999).

\bibitem{Mangiameli} Mangiameli, P., Chen, S. K.  West, D. A comparison of SOM neural network and hierarchical clustering methods. Eur. J. Oper. Res. \textbf{93}, 402–417 (1996).

\bibitem{Herrero} Herrero, J., Valencia, A. and Dopazo, J. A hierarchical unsupervised growing neural network for clustering gene expression patterns. Bioinformatics \textbf{17}, 126–136 (2001).

\bibitem{Okamoto} Okamoto, M., Bu, N., and Tsuji, T. Unsupervised learning for hierarchical clustering using statistical information. in Advances in Neural Networks 834–839 (2004).

\bibitem{Shahid} Nazish Shahid, Comparison of hierarchical clustering and neural network clustering: an analysis on precision dominance, Sci. Rep. \textbf{13}, 5661 (2023).

\bibitem{1} D. A. Goss and R. W. West, "Introduction to the Optics of the Eye" (Butterworth-Hienemann, 2001). 
\bibitem{2} M. P. Keating, Geometric, "Physical and Visual Optics" (Butterworth-Hienemann, 2002). 
\bibitem{3} W. Tasman and E. A. Jaeger, "Duane's Ophtalmology" (LLW, 2013). 
\bibitem{4} T. Callina and T. P. Reynolds, "Traditional methods for the treatment of presbyopia: spectacles, contact lenses, bifocal contact lenses", Ophthalmol. Clin. North Am. \textbf{19}, 25-33 (2006). 
\bibitem{5} Patent Application WO2006011937A2, ``Fluidic Adaptive Lens''. 
\bibitem{6} N. Hazan, A. Banerjee, H. Kim, C. Mastrangelo, ``Tunable-focus lens for adaptive eyeglasses'', Opt. Express \textbf{25}, 1221 (2017).
\bibitem{7} O. Takayama, F. Minotti, G. Puentes, ``Tunable Fluidic Lenses with High Dioptric Power'', OSA Continuum \textbf{1}, 181 (2018).
\bibitem{8} G. Puentes, D. Voigt, A. Aiello and J. P. Woerdman, "Experimental observation of depolarized light scattering", Opt. Lett. \textbf{30}, 3216-3219 (2005). 
\bibitem{9} G. Puentes and F. Minotti, Spectral Characterization of Optical Aberrations in Fluidic Lenses,  \textbf{11}, 1299393. (2024). 
\bibitem{10} G. Puentes Invention Disclosure Nr.20170102760. ``Adaptive Fluidic Lenses for Subnormal Vision Segment'' (AR109794B1). 
\bibitem{10b} G. Puentes, A. Datta, A. Feito, J. Eisert, M. B. Plenio, I. A. Walmsley, "Entanglement quantification from incomplete measurements: applications using photon-number-resolving weak homodyne detectors", New Journal of Physics \textbf{12}, 033042 (2010). 
\bibitem{11} F. Schneider, J. Draheim, R. Kamberger, U. Wallrabe, Process and material properties of polydimethylsiloxane (PDMS) for Optical MEMS, Sensor and Actuators A \textbf{151}, 95-99 (2006).
\bibitem{12} M. Vallet, B. Berge, and L. Vovelle, "Electrowetting of water and aqueos solutions on poly-ethilene-terephthalate insulating films", Polymer \textbf{37}, 2465-2470 (1996). 
\bibitem{13} T. Krupenking, S. Yang, P. Mach, "Tunable liquid microlens", Appl. Phys. Lett. \textbf{82}, 316-318 (2003). 
\bibitem{14} S. Kuiper andB. H. Hendriks, "Variable-focus liquid lens for miniature cameras", Appl. Phys. Lett. \textbf{85}, 1128-1130 (2004). 
\bibitem{15} G. C. Knollman, J. L. Bellini, J. L. Weaver, "Variable-focus liquid-filled hydroacoustic lens",  J. Acoust. Soc. Am. \textbf{49}, 253-261 (1971).
\bibitem{16} N. Sigiura and S. Morita, "Variable-docus liquid-filled optics lens", App. Opt. \textbf{32}, 4181-4186 (1993). 
\bibitem{17} D. Y. Zhang, V. Lien, Y. Berdichevsky, J. Choi, and Y. H. Lo, "Fluidic adaptivelens with high focal length tenability", App. Phys. Lett. \textbf{82}, 3171-3172 (2003). 
\bibitem{18}K. H. Joeng, G. L. Liu, N. Chronis, And L. P. Lee, "Tunable microdoublets lens array", Opt. Express \textbf{12}, 2494-2500 (2004). 
\bibitem{19} J. Chen, W. Wang, J. Fang, and K. Varahramyan, "Variable focusing microlens with microfluidic chip", J. Micromech. Microeng. \textbf{14}. 675-680 (2004). 
\bibitem{20} N. Chronis, G. L. Liu, K. H. Jeong, and L. P. Lee, "Tunable liquid filled micro-lens array integrated with microfludidic network", Opt. Express \textbf{11}, 2370-2378 (2003). 
\bibitem{21} P. M. Moran, s. Dharmatilleke, A. H. Khaw, and K. W. Tan, ``Fluid lenses with variable focal length'', App. Phys. Lett. \textbf{88}, 041120 (2006). 
\bibitem{22} H. Ren and S-T Wu, ``Variable-focus liquid lens'', Opt. Express \textbf{15}, 5931-5936 (2007).
\bibitem{23} N. A. Polson, and M. A. Hayes, ``Microfluidics controllling fluids in small places'', Anal. Chem. \textbf{73}, 312A-319A (2001).

\bibitem{37} Siemon, M. S. N., Shihavuddin, A. S. M.  Ravn-Haren, G. Sequential transfer learning based on hierarchical clustering for improved performance in deep learning based food segmentation. Sci. Rep. \textbf{11}, 813 (2021).

\bibitem{38} Maione, C., Nelson, D. R.  Barbosa, R. M. Research on social data by means of cluster analysis. Appl. Comput. Inform. \textbf{15}, 153–162 (2008).

\end{thebibliography}
\end{document}